\documentclass[twocolumn,showpacs,amsmath,amssymb,10pt]{revtex4}
\usepackage{graphicx,color}
\usepackage{amsmath}
\usepackage{amssymb}

\usepackage{amsfonts}
\usepackage{bm}
\newcommand{\ra}{{\, \rightarrow\, }}

\newcommand{\ot}{{\,\otimes\,}}
\newcommand{{\Cd}}{{\mathbb{C}^d}}

\newcommand{\tr}{\mathrm{Tr}}

\def\oper{{\mathchoice{\rm 1\mskip-4mu l}{\rm 1\mskip-4mu l}%
{\rm 1\mskip-4.5mu l}{\rm 1\mskip-5mu l}}}
\def\<{\langle}
\def\>{\rangle}

\newtheorem{definition}{Definition}

\newtheorem{Example}{Example}

\begin{document}
\title{\textbf{Witnessing non-Markovianity of quantum evolution}}
\author{Dariusz Chru\'sci\'nski and Andrzej Kossakowski}
\affiliation{Institute of Physics, Nicolaus Copernicus University \\
Grudzi\c{a}dzka 5/7, 87--100 Toru\'n, Poland}

\begin{abstract}

We provide further characterization of non-Markovian quantum dynamics based on the concept of divisible dynamical maps. In analogy to entanglement witness  we propose a {\em non-Markovianity witness} and introduce the corresponding  measure of non-Markovianity. We also provide characterization of non-Markovianity in terms of entropic quantities, fidelity and Wigner-Yanase-Dyson skew information.

\end{abstract}


\maketitle

\section{Introduction}

The dynamics of open quantum
systems attracts nowadays considerable attention.
It is relevant not only for a better understanding of
quantum theory but it is fundamental in various modern
applications of quantum mechanics. Since the system environment
interaction causes dissipation, decay and
decoherence it is clear that the dynamics of open
systems is fundamental in modern quantum technologies,
such as quantum communication, cryptography,
computation and quantum metrology.

The traditional approach to the
dynamics of an open quantum system consists in applying a suitable Born-Markov approximation leading to the celebrated
quantum Markov semigroup \cite{GKS,Lindblad} which neglects all
memory effects. Recent theoretical activity and
technological progress  show the importance of more refine approach based on
non-Markovian evolution. Non-Markovian quantum dynamics becomes in recent years very active field of both theoretical and experimental research and there are a lot of papers devoted to this topic ( see e.g. \cite{Strunz1}--\cite{B-review} and references therein).

Surprisingly,  the concept of (non)Markovianity is not
uniquely defined. One approach is based on the idea of the
composition law which is essentially equivalent to the idea of
divisibility \cite{Wolf2}. A dynamical map $\Lambda_t$ is divisible if $\Lambda_t = V_{t,s} \Lambda_s$ and $V_{t,s}$ is completely positive and trace preserving for all $t\geq s$, that is, it gives rise to 2-parameter family of legitimate propagators. The essential property of $V_{t,s}$ is the following (inhomogeneus) composition law
\begin{equation}\label{c-law}
    V_{t,s} \, V_{s,u} = V_{t,u} \ ,
\end{equation}
for all $t\geq s\geq u$. It is clear that (\ref{c-law}) generalizes semigroup property.
This approach was used  by Rivas,
Huelga and Plenio (RHP) \cite{RHP} to construct the corresponding
measure of non-Markovianity which measures the deviation from
divisibility. In this paper we assume that Markovian dynamics is represented by divisible dynamical map.
It should be stressed that Markovian dynamics (divisible map) is entirely characterized by the properties of the local in time generators $L_t$, that is, if $\Lambda_t$ satisfies $\dot{\Lambda}_t =L_t \Lambda_t$, then $\Lambda_t$ corresponds to Markovian dynamics if and only if $L_t$ has the standard form \cite{GKS,Lindblad} for all $t\geq 0$, that is,
\begin{equation*}\label{}
    L_t \rho = -i[H_t,\rho]  + \sum_\alpha \left( V_\alpha(t) \rho V_\alpha^\dagger(t) - \frac 12 \{ V_\alpha^\dagger(t)V_\alpha(t),\rho\} \right) \ ,
\end{equation*}
with time dependent Hamiltonian $H_t$ and noise operators $V_\alpha(t)$.

A different approach is advocated by Breuer, Laine and
Piilo (BLP) in Ref. \cite{BLP}. BLP define non-Markovian dynamics as
a time evolution for the open system characterized by a temporary
flow of information from the environment back into the system and
manifests itself as an increase in the distinguishability of pairs
of evolving quantum states:
\begin{equation}\label{BLP-c}
    \sigma(\rho_1,\rho_2;t) = \frac 12 \frac{d}{dt} \, ||\Lambda_t(\rho_1-\rho_2)||_1\ ,
\end{equation}
where $||A||_1 = {\rm Tr}\sqrt{A^\dagger A}$ denotes the trace norm. According to \cite{BLP} the dynamics $\Lambda_t$ is markovian iff $ \sigma(\rho_1,\rho_2;t) \leq 0$ for all pairs of states $\rho_1,\rho_2$ and $t \geq 0$. Optimizing over $\rho_1,\rho_2$ enables to construct suitable non-Markovianity measure \cite{BLP} (see the recent paper \cite{optimal} discussing the properties of the optimal pair $\rho_1,\rho_2$). It turns out that condition $\sigma(\rho_1,\rho_2;t) \leq 0$ is less restrictive than requirement of complete positivity for $V_{t,s}$ and one can construct  $\Lambda_t$ which is non-Markovian (not divisible) but still  gives rise to the negative flow of information (see \cite{versus1,versus2,versus3}). Other measures of non-Markovianity based on quantum Fisher information \cite{inne1}, fidelity \cite{inne2}, departure from divisibility \cite{inne3} and quantum mutual information \cite{inne4}, were proposed as well.


In the present paper we provide further characterization of non-Markovian quantum dynamics. We stress in our approach Markovianity of evolution corresponds to the divisibility of the corresponding dynamical map. In the next section we propose a {\em non-Markovianity witness}. In analogy to entanglement witness non-Markovianity witness is defined by Hermitian not positive operator in $\mathcal{H} \ot \mathcal{H}$. Using this concept we introduce a measure of non-Markovianity which is essentially equivalent to the RHP measure. We also provide characterization of non-Markovianity in terms of entropic quantities, fidelity and Wigner-Yanase-Dyson skew information.  Simple examples illustrate the differences and similarities between these characteristics. In particular we provide another example of quantum evolution which is non-Markovian (not divisible) but still satisfies BLP criterion (\ref{BLP-c}). Final conclusions are collected in the last section.

\section{Non-Markovianity witness}

Let us recall, that if $\mathcal{E} : \mathcal{T}(\mathcal{H})\rightarrow  \mathcal{T}(\mathcal{H})$ is a linear trace preserving map, then $\mathcal{E} $ is a quantum channel   if and only if
\begin{equation}\label{contr_0}
    || (\oper \ot \mathcal{E})X ||_1 \leq ||X||_1\ ,
\end{equation}
for all  $X=X^\dagger \in \mathcal{B}(\mathcal{H}\ot \mathcal{H})$. Traditionally $\mathcal{T}(\mathcal{H})$ denotes the vector space of trace class operators, i.e. $x \in \mathcal{T}(\mathcal{H})$ if $||x||_1 = {\rm Tr}\sqrt{xx^\dagger} < \infty$. It is clear that if $\mathcal{H}$ is finite dimensional, then $\mathcal{T}(\mathcal{H})= \mathcal{B}(\mathcal{H})$.
Actually, $\oper \ot \mathcal{E}$ is a contraction for all $X$ not necessarily hermitian. Hence, a dynamical map $\Lambda_t$ is Markovian
if and only if
\begin{equation}\label{contr}
    \lambda_t(X) := \frac {d}{dt}\,  || (\oper \ot \Lambda_t)X ||_1 \leq 0\ .
\end{equation}

\begin{definition}
We call $X^\dagger = X$ non-Markovianity witness for $\Lambda_t$ iff $\lambda_t(X)>0$ for some $t > 0$.
\end{definition}
Note, that if $X \geq 0$ then $(\oper \ot \Lambda_t)X \geq 0$ and hence $|| (\oper \ot \Lambda_t)X ||_1 = {\rm Tr} X$ which implies $\lambda_t(X) =0$. Therefore, a necessary condition for $X$ to be non-Markovianity witness is $X \ngeq 0$. Recall, that it is also a necessary condition for an entanglement witness.

We can propose a natural measure of non-Markovianity
\begin{equation}\label{M}
    \mathcal{N}[\Lambda_t] = \sup_{||X||_1=1} \, \int_{ \lambda_t(X) > 0 }  \lambda_t(X) \, dt \ ,
\end{equation}
that is, the formula (\ref{M}) choses the {\em optimal} witness and measures the violation of $\lambda_t(X) \leq 0$ along the trajectory.

Note, that condition (\ref{contr}) implies
\begin{equation}\label{contr-1}
\frac {d}{dt}\,    || \Lambda_t x ||_1 \leq 0\ ,
\end{equation}
for all  $x=x^\dagger \in \mathcal{B}(\mathcal{H})$, and (\ref{contr-1}) implies BLP condition
\begin{equation}\label{contr-2}
\frac {d}{dt}\,    || \Lambda_t (\rho-\sigma) ||_1 \leq 0\ ,
\end{equation}
for all density operators $\rho$ and $\sigma$ in $\mathcal{H}$. It is, therefore clear, that BLP definition of non-Markovianity is less restrictive.

\begin{Example} \label{EX1}  {\em
Consider pure decoherence of a qubit system described by the following local generator
\begin{equation}\label{}
    L_t \rho = \frac 12 \gamma(t) (\sigma_z \rho \sigma_z - \rho) \ ,
\end{equation}
The corresponding evolution of the density matrix reads
\begin{equation}\label{}
    \rho_t = \left( \begin{array}{cc} \rho_{11} & \rho_{12} e^{-\Gamma(t)} \\ \rho_{12} e^{-\Gamma(t)} & \rho_{22} \end{array} \right) \ ,
\end{equation}
where $\Gamma(t) = \int_0^t \gamma(\tau) d\tau$. The evolution is Markovian  iff $\gamma_t \geq 0$.
Taking $X_0 = \frac 12 \sigma_x \ot \sigma_x$ one finds $\lambda_t(X_0) = - \gamma(t) e^{-\Gamma(t)}$. Hence $\lambda_t(X_0) > 0$ whenever $\gamma(t) < 0$. We claim that $X_0$ optimizes (\ref{M}). Let us recall, that RHP compute
\begin{equation}\label{}
    g(t) = \lim_{\epsilon \ra 0+} \frac{|| P^+ + \epsilon (\oper \ot L_t)P^+||_1 - 1}{\epsilon}\ ,
\end{equation}
and get $g(t) = |\gamma(t)|$ whenever $\gamma(t) < 0$. One has
\begin{equation}\label{}
    \mathcal{N}_{\rm RHP}[\Lambda_t] = \sum_k \Delta \Gamma_k \ ,
\end{equation}
where $\Delta \Gamma_k = |\Gamma(t_k + \Delta_k) - \Gamma(t_k)|$ and $\gamma(t) < 0$ for $t \in (t_k,t_k+\Delta_k)$. Similarly, one finds
\begin{equation}\label{}
    \mathcal{N}[\Lambda_t] = \sum_k |  e^{-\Gamma(t_k + \Delta_k)} - e^{-\Gamma(t_k)}|   \ .
\end{equation}
It is clear that $\mathcal{N}[\Lambda_t]>0$ if and only if $\mathcal{N}_{\rm RHP}[\Lambda_t]>0$.

}
\end{Example}

Suppose that a dynamical map $\Lambda_t$ possess time-independent eigenvector, that is, $\Lambda_t f = \mu_t f$, where $\mu_t$ belongs to the unit disc in the complex plane.  Markovianity implies that $\frac{d}{dt}|\mu_t| \leq 0$ for all $t\geq 0$. There are many examples of quantum dynamics where the off-diagonal elements behave according to
\begin{equation}\label{}
    \Lambda_t(|i\>\<j|) = G_{ij}(t) |i\>\<j|\ , \ \ \ i \neq j \ .
\end{equation}
Hence, if $\frac{d}{dt}|G_{ij}(t)| \nleqslant 0$ for at least one pair $(ij)$, then $\Lambda_t$ is non-Markovian.

Note, that if all eigenvectors $f_\alpha$ of $\Lambda_t$ are time independent
\begin{equation}\label{}
    \Lambda_t f_\alpha = \mu_\alpha(t) f_\alpha\ ,\ \ \ \alpha=1,\ldots,({\rm dim}\mathcal{H})^2\ ,
\end{equation}
then
$\Lambda_t \Lambda_u = \Lambda_u \Lambda_t$, for all $t,u \geq 0$. One may call it commutative dynamics.
 If the dynamics is Markovian, then
\begin{equation}\label{}
    \frac{d}{dt}\, ||\Lambda_t f_\alpha||_1 =  \frac{d}{dt}\, |{\mu}_\alpha(t)|\, ||f_\alpha||_1 \leq 0 \ .
\end{equation}
Therefore, for a class of commutative dynamical maps Markovianity implies monotonicity of all $|{\mu}_\alpha(t)|$. Actually, pure decoherence dynamics belongs to this class. One has
\begin{equation*}
    \Lambda_t (|k\>\<k|)=|k\>\<k|\ (k=1,2)\ , \ \   \Lambda_t (|1\>\<2|)=  e^{-\Gamma(t)}|1\>\<2|\ ,
\end{equation*}
and hence $\mu_1(t)=\mu_2(t)=1$ and $\mu_3(t)=\mu_4(t) = e^{-\Gamma(t)}$.
Hence Markovianity implies $\gamma(t) \geq 0$.

\begin{Example} \label{EX2}  {\em
Consider the dynamics governed by the local in time generator
\begin{equation}\label{}
    L_t \rho = \gamma(t) \left( \omega_t \, {\rm Tr}\, \rho - \rho \right) \ ,
\end{equation}
where $\omega_t$ is a family of Hermitian operators satisfying ${\rm Tr}\,\omega_t = 1$.
The above generator gives rise to Markovian evolution iff $L_t$ has the standard form \cite{GKS,Lindblad} for all $t\geq 0$, that is, iff $\gamma(t) \geq 0$ and $\omega_t$ defines a legitimate state, i.e. $\omega_t \geq 0$. The corresponding solution of the Master equation $\dot{\rho}_t = L_t \rho_t$ with an initial condition $\rho_{t=0} = \rho$ reads as follows
\begin{equation}\label{}
    \rho_t = e^{-\Gamma(t)} \rho + [1- e^{-\Gamma(t)}] \Omega_t\, {\rm Tr}\rho\ ,
\end{equation}
where $\Omega_t = \frac{1}{e^{\Gamma(t)}-1} \int_0^t \gamma(\tau) e^{\Gamma(\tau)} \, \omega_\tau d\tau $ (note, that if $\rho$ is density matrix then ${\rm Tr}\rho=1$).  It is therefore clear
that $L_t$ generates a legitimate quantum evolution iff $\Gamma(t) \geq 0$ and $\Omega(t) \geq 0$, that is, $\Omega_t$ defines a legitimate state (note, that ${\rm Tr}\, \Omega_t=1$). In particular, if $\omega_t = \omega$ is time independent, then $\Omega_t = \omega$ and the solution simplifies to  a convex combination of the initial state $\rho$ and the asymptotic invariant state $\omega$: $\rho_t = e^{-\Gamma(t)} \rho + [1- e^{-\Gamma(t)}]\, \omega$. One easily shows that the evolution is Markovian iff $\gamma(t) \geq 0$ and $\omega_t$ is a legitimate density operator (that is, $\omega_t \geq 0$). Consider now the BLP condition (\ref{BLP-c}). One has $\rho_t - \sigma_t = e^{-\Gamma(t)} (\rho-\sigma)$ and hence
\begin{equation*}
    \frac{d}{dt}\, ||\rho_t - \sigma_t||_1  = - \gamma(t)\, e^{-\Gamma(t)} ||\rho-\sigma||_1 \leq 0 \ ,
\end{equation*}
implies only $\gamma(t) \geq 0$ but says nothing about $\omega_t$. It shows that any $\omega_t$ which gives rise to $\Omega_t \geq 0$ leads to the evolution satisfying condition (\ref{BLP-c}) but only $\omega_t \geq 0$ gives rise to Markovian dynamics. Hence, we may have non-Markovian  dynamics ($\omega_t \ngeq 0$ but $\Omega_t \geq 0$) which satisfies BLP condition (\ref{BLP-c}) for all $t\geq 0$. Clearly, such non-Markovian dynamics has vanishing BLP non-Markovianity measure.
}
\end{Example}



\section{Entropic witnesses}

Let us recall that the relative entropy defined by
\begin{equation}\label{}
    S(\rho\,||\,\sigma) = {\rm Tr}(\rho [\log \rho - \log \sigma] )\ .
\end{equation}
(one assumes that $S(\rho\,||\,\sigma) = \infty$ when supports of $\rho$ and $\sigma$ do not satisfy ${\rm supp}\, \rho \subset {\rm supp}\,\sigma$) enjoys
\begin{equation}\label{}
    S(\mathcal{E}(\rho)\,||\,\mathcal{E}(\sigma)) \leq  S(\rho\,||\,\sigma) \ ,
\end{equation}
for any quantum channel $\mathcal{E}$. Therefore, if $\Lambda_t$ is a dynamical map, then
$  S(\Lambda_t(\rho)\,||\,\Lambda_t(\sigma)) \leq  S(\rho\,||\,\sigma)$ for any $t \geq 0$. Hence, if $\Lambda_t$ corresponds too Markovian evolution, then
\begin{equation}\label{S1}
    \frac{d}{dt}\, S(\Lambda_t(\rho)\,||\,\Lambda_t(\sigma)) \leq 0\ ,
\end{equation}
for each pair of initial states $\rho$ and $\sigma$. Note, that if $\sigma_0$ is an invariant state, i.e. $\Lambda_t(\sigma_0) = \sigma_0$, then (\ref{S1}) simplifies to
\begin{equation}\label{S1a}
    \frac{d}{dt}\, S(\Lambda_t(\rho)\,||\,\sigma_0) \leq 0\ ,
\end{equation}
for each $\rho$. In particular, if $\sigma_0= \mathbb{I}/n$ is maximally mixed, then $S(\rho\,||\,\sigma_0) = \log n - S(\rho)$, and hence the formula (\ref{S1a}) reduces to
\begin{equation}\label{S1b}
    \frac{d}{dt}\, S(\Lambda_t(\rho)) \geq 0\ ,
\end{equation}
that is, the von Neumann entropy monotonically increases for each initial state $\rho$.

Consider once more pure decoherence of a qubit form Example \ref{EX1}.
Note, that $L_t \mathbb{I} = 0$ and hence the maximally mixed state is invariant. Therefore, Markovianity implies (\ref{S1b}).
Now, for the 2-level system
\begin{equation}\label{}
    \frac{d}{dt}\, S(\rho_t) = - \dot{\lambda}^+_t \log\frac{\lambda^+_t}{\lambda^-_t}\ ,
\end{equation}
where $\lambda^+_t \geq \lambda^-_t$ are eigenvalues of $\rho_t$. Hence $\frac{d}{dt}\, S(\rho_t) \geq 0$ if $\dot{\lambda}^+_t \leq 0$.
One easily finds
\begin{equation*}\label{}
    \lambda^\pm_t = \frac 12 \left( 1 \pm \sqrt{ (\rho_{11}-\rho_{22})^2 + |\rho_{12}|^2 e^{-2\Gamma(t)} } \right) \ .
\end{equation*}
It is therefore clear that $S(\rho_t)$ monotonically increases if and only if $\dot{\Gamma}(t) = \gamma(t) \geq 0$.

The above scheme may be immediately repeated for the well known families of generalized  Renyi $S_\alpha$ and Tsallis $T_q$ relative entropies
\begin{equation}\label{}
    S_\alpha(\rho\, ||\, \sigma) = \frac{1}{\alpha-1} \log\Big[ {\rm Tr}\,\rho^\alpha \sigma^{1-\alpha} \Big] \ ,
\end{equation}
for $\alpha \in [0,1) \cup (1,\infty)$, and
\begin{equation}\label{}
    T_q(\rho\, ||\, \sigma) = \frac{1}{1-q}\, \Big[ 1 - {\rm Tr} \,\rho^q \sigma^{1-q} \Big] \ ,
\end{equation}
for $q\in [0,1)$. Note, that in the limit
$$ \lim_{\alpha \ra 1} S_\alpha(\rho\, ||\, \sigma) =  \lim_{q \ra 1} T_q(\rho\, ||\, \sigma) = S(\rho\, ||\, \sigma)\ , $$
one recovers von Neumann relative entropy. It turns out \cite{QIT,Petz-08} that if $\mathcal{E}$ is a quantum channel then $S_\alpha$
satisfies $S_\alpha(\mathcal{E}(\rho)\, ||\, \mathcal{E}(\sigma)) \leq  S_\alpha(\rho\, ||\, \sigma)$ for $\alpha \in [0,1) \cup (1,2]$  and the same applies for $T_q$ for $q \in [0,1)$. If $\Lambda_t$ is a divisible map, then
\begin{equation}\label{ST1}
    \frac{d}{dt}\, S_\alpha(\Lambda_t\rho\, ||\, \Lambda_t\sigma)  \leq 0 \ , \ \ \ \
    \frac{d}{dt}\, T_q(\Lambda_t\rho\, ||\, \Lambda_t\sigma)  \leq 0 \ ,
\end{equation}
for $\alpha \in [0,1) \cup (1,2]$ and  $q\in [0,1)$. Again, if the maximally mixed state is invariant, then (\ref{ST1}) gives rise to
\begin{equation}\label{ST1b}
    \frac{d}{dt}\, S_\alpha(\Lambda_t(\rho)) \geq 0\ ,\ \ \ \  \frac{d}{dt}\, T_q(\Lambda_t(\rho)) \geq 0\ ,
\end{equation}
which generalize (\ref{S1b}).

\section{Fidelity witness}

Given two density operators $\rho$ and $\sigma$ one defines Uhlmann fidelity
\begin{equation}\label{}
    F(\rho,\sigma) = \Big({\rm Tr}\, \Big[ \sqrt{\sqrt{\rho}\, \sigma\, \sqrt{\rho}} \Big] \Big)^2\ .
\end{equation}
Equivalently, one has $ F(\rho,\sigma) = || \sqrt{\rho} \sqrt{\sigma} ||_1^2$ which shows that $F(\rho,\sigma) = F(\sigma,\rho)$.
One proves
\begin{equation}\label{}
    1 - F(\rho,\sigma) \leq D[\rho,\sigma] \leq \sqrt{1-F(\rho,\sigma)^2}\ .
\end{equation}
Moreover, if $\mathcal{E}$ is a quantum channel, then $F(\mathcal{E}(\rho),\mathcal{E}(\sigma)) \geq F(\rho,\sigma)$ which implies that for the Markovian evolution one has
\begin{equation}\label{F}
    \frac{d}{dt} F(\Lambda_t(\rho),\Lambda_t(\sigma)) \geq 0 \ .
\end{equation}
Again, if $\sigma_0$ is an invariant state, then Markovianity implies $  \frac{d}{dt} || \sqrt{\rho_t}\sqrt{\sigma_0}||_1 \geq 0$. In particular, if $\sigma_0$ is maximally mixed, then
\begin{equation}\label{}
    \frac{d}{dt} || \sqrt{\rho_t}||_1 = \frac{d}{dt} \, {\rm Tr} \sqrt{\rho_t}\,  \geq 0\ .
\end{equation}
Note, that the above condition is equivalent to
\begin{equation}\label{}
\frac{d}{dt} T_{\frac 12}(\rho_t) \geq 0\ .
\end{equation}
If $\sigma_0=|\psi_0\>\<\psi_0|$ is pure, then (\ref{F}) reduces to a very simple condition
\begin{equation}\label{}
    \frac{d}{dt} \< \psi_0|\rho_t|\psi_0\> \geq 0 \ ,
\end{equation}
for all $t\geq 0$. It means that the overlap of $\rho_t$ with an invariant vector state $|\psi_0\>$ monotonically increases.

\begin{Example}[Spin-boson model] {\em
Consider once more the evolution of a qubit system described by the following local generator
\begin{equation}\label{}
    L_t \rho = - \frac{is(t)}{2} [\sigma_+\sigma_-,\rho] + \gamma(t) (\sigma_- \rho \sigma_+ - \frac 12 \{\sigma_+\sigma_-,\rho\} ) \ ,
\end{equation}
where $s(t) = - 2{\rm Im}\, \frac{\dot{G}(t)}{G(t)}$ and  $\gamma(t) = - 2{\rm Re}\, \frac{\dot{G}(t)}{G(t)}$, and the function $G(t)$ satisfies non-local equation
\begin{equation*}
    \dot{G}(t) = - \int_0^t  f(t-\tau)\, G(\tau)\, d\tau\ , \ \ \ G(0) = 1\ ,
\end{equation*}
with $f(t)$ being a correlation function of the (bosonic) reservoir. The standard raising and lowering operators read: $\sigma_+=|2\>\<1|$ and $\sigma_-=|1\>\<2|$. The corresponding evolution is given by the following formulae
\begin{equation*}\label{}
    \rho_{11}(t) = \rho_{11} + (1- |G(t)|^2)\rho_{22} \ , \ \ \rho_{22}(t) = |G(t)|^2 \rho_{22} \ ,
\end{equation*}
and the off-diagonal element $ \rho_{12}(t) = G^*(t) \rho_{12}$. It is clear that the ground state $|\psi_0\> = |1\>$ defines an invariant state, and hence Markovianity implies
\begin{equation}\label{}
    \frac{d}{dt} \< \psi_0|\rho_t|\psi_0\> = \frac{d}{dt} \rho_{11}(t) \geq 0 \ ,
\end{equation}
which is equivalent to $\frac{d}{dt}\,  |G(t)|^2 \leq 0$ and hence to $\gamma(t)\geq 0$.
}
\end{Example}

\section{Wigner-Yanase-Dyson skew information}

Consider the following quantity
\begin{equation}\label{}
    I(\rho,X) = - \frac 12 {\rm Tr}\, [\sqrt{\rho},X]^2\ ,
\end{equation}
where $X^\dagger = X$ is an observable, introduced by Wigner  and Yanase \cite{WY}. One finds
$I(\rho,X) =  {\rm Tr}[ \rho X^2 - \sqrt{\rho} X \sqrt{\rho} X ]$ and hence if $\rho$ is pure then $I(\rho,X)$ reduces to the variance
$V(\rho,X) =  {\rm Tr}(\rho X^2) - {\rm Tr}(\rho X)^2$. This quantity was generalized by Dyson
\begin{equation}\label{WYD}
    I_p(\rho,X) = - \frac 12 {\rm Tr}\, [\rho^p,X][\rho^{1-p},X]\ ,
\end{equation}
for arbitrary $0< p <1$, and its is called Wigner-Yanase-Dyson skew information. The convexity of $I=I_{1/2}$ was already proved by Wigner and Yanase \cite{WY}, and for the general case $p\in (0,1)$ -- the celebrated Wigner-Yanase-Dyson conjecture -- it was proved by Lieb \cite{Lieb}.

It was proved by Petz \cite{Petz-08,Ohya-Petz} that if $\mathcal{E}$ is a quantum channel, then
\begin{equation}\label{}
    I_p(\mathcal{E}(\rho),X) \geq I_p(\rho,\mathcal{E}^*(X))\ ,
\end{equation}
where $\mathcal{E}^*$ denotes a dual channel (Heisenberg picture) defined by ${\rm Tr}(\mathcal{E}(\rho) X) = {\rm Tr}(\rho \mathcal{E}^*(X))$. Hence, if $\sigma_0$ is an invariant state of $\Lambda_t$, then
\begin{equation}\label{}
    I_p(\sigma_0,X) \geq I_p(\sigma_0,\Lambda_t^*(X))\ .
\end{equation}
Similarly, if $X_0$ is an invariant observable (constant of motion), i.e. $\Lambda^*_t(X_0) = X_0$, then
\begin{equation}\label{}
    I_p(\Lambda_t(\rho),X_0) \geq I_p(\rho,X_0)\ .
\end{equation}
In conclusion, if $\Lambda_t$ is Markovian, then
\begin{equation}\label{WYD-1}
     \frac{d}{dt}\, I_p(\sigma_0,\Lambda_t^*(X))\leq 0\ ,
\end{equation}
and
\begin{equation}\label{WYD-2}
    \frac{d}{dt}\, I_p(\Lambda_t(\rho),X_0) \geq 0 \ .
\end{equation}
Interestingly, if $\sigma_0 = |\psi_0\>\<\psi_0|$ is a pure state, then Markovianity implies
\begin{equation}\label{WYD-pure}
     \frac{d}{dt}\, \Big[ \<\psi_0| X_t^2|\psi_0\> - \<\psi_0|X_t|\psi_0\>^2 \Big] \leq 0 \ ,
\end{equation}
where $X_t = \Lambda_t^*(X)$. It shows that for Markovian dynamics dispersion in the invariant pure state monotonically decreases.

Consider spin-boson model defined in Example 2. One finds for the evolution in the Heisenberg picture
\begin{equation*}\label{}
    X_{11}(t) = X_{11} \ , \ \ X_{22}(t) = (1-|G(t)|^2) X_{11} + |G(t)|^2 X_{22} \ ,
\end{equation*}
and the off-diagonal element $ X_{12}(t) = G(t)X_{12}$. Taking as an invariant state $\sigma_0=|1\>\<1|$ the formula
(\ref{WYD-pure}) implies
\begin{equation}\label{}
    \frac{d}{dt}\, \left[ X_{11} + |G(t)|^2 |X_{12}|^2 \right] \leq 0 \ ,
\end{equation}
which is equivalent to $\frac{d}{dt}\,  |G(t)|^2 \leq 0$ and hence to $\gamma(t)\geq 0$.

\section{Conclusions}

We have provided several criteria enabling one to witness the non-Markovianity of quantum evolution. Note, that passing to the Heisenberg picture (that is, using the dual map $\Lambda_t^*$) one may reformulate the formula (\ref{contr}) as follows: $\Lambda_t^*$ corresponds to the Markovian dynamics in the Heisenberg picture iff
\begin{equation}\label{contr-H}
    \widetilde{\lambda}_t(X) := \frac {d}{dt}\,  || (\oper \ot \Lambda^*_t)X || \leq 0\ ,
\end{equation}
where $|| A ||$ denotes the operator norm in $\mathcal{B}(\mathcal{H}\ot \mathcal{H})$. Both criteria are perfectly equivalent:
${\lambda}_t(A) \leq 0$ for all $A \in \mathcal{T}(\mathcal{H}  \ot \mathcal{H})$ and $t \geq 0$ if and only if $\widetilde{\lambda}_t(B) \leq 0$ for all $B \in \mathcal{B}(\mathcal{H}  \ot \mathcal{H})$ and $t \geq 0$. Interestingly, Wigner-Yanase-Dyson skew information merges both pictures and provides the constraints for Markovianity of quantum evolution given by  (\ref{WYD-1}) and (\ref{WYD-2}). It is clear that the witnesses of non-Markovianity presented in this paper may be immediately generalized.

One may for example try to use generalized entropic measures \cite{Ros}
$S_f(\rho) = {\rm Tr}f(\rho)$, where  $f(p)$ is a smooth strictly concave real function defined
for $p \in [0,1]$ satisfying $f(0) = f(1) = 0$. They provide further generalization of Renyi and Tsallis entropies.

Note, that if $F$ is a function satisfying monotonicity condition $F(\mathcal{E}(\rho)) \leq F(\rho)$ for any quantum channel $\mathcal{E}$, then Markovianity of $\Lambda_t$ implies $\frac{d}{dt} F(\Lambda_t(\rho)) \leq 0$ for all $t \geq 0$. Similarly, if $\widetilde{F}([\oper_A \ot \mathcal{E}](\rho_{AB})) \leq \widetilde{F}(\rho_{AB})$, then $\frac{d}{dt} \widetilde{F}([\oper_A \ot \Lambda_t](\rho_{AB})) \leq 0$ for Markovian evolution. This property was used by Rivas et. al. \cite{RHP} by taking as $\widetilde{F}$ genuine entanglement measure. Similarly  Luo et. al. \cite{inne4}  take as $\widetilde{F}$ the mutual information.
Fidelity and relative entropy may be replaced by any function $G$  satisfying monotonicity condition $G(\mathcal{E}(\rho),\mathcal{E}(\sigma)) \leq G(\rho,\sigma)$. Markovianity of $\Lambda_t$ implies $\frac{d}{dt} G(\Lambda_t(\rho),\Lambda_t(\sigma)) \leq 0$ for all $t \geq 0$.

\acknowledgements
This work was partially supported by the National Science Center project  
DEC-2011/03/B/ST2/00136.

\end{document}